\documentclass[12pt,a4paper]{article}
\usepackage{a4wide}
\usepackage{latexsym}
\usepackage{epsf}
\usepackage{amssymb}
\usepackage{amsbsy}
\usepackage{amscd}
\usepackage{amsmath}
\usepackage{amstext}
\usepackage{graphicx}
\begin{document}
\def\be{\begin{equation}}
\def\ee{\end{equation}}
\def\bea{\begin{eqnarray}}
\def\eea{\end{eqnarray}}
\def\bi{\begin{itemize}}
\def\ei{\end{itemize}}
\def\pd{\partial}
\def\a{\alpha}
\def\b{\beta}
\def\g{\gamma}
\def\d{\delta}
\def\m{\mu}
\def\n{\nu}
\def \h{\mathcal{H}}
\def \hh{\mathcal{G}}
\def\t{\tau}
\def\p{\pi}
\def\th{\theta}
\def\l{\lambda}
\def\O{\Omega}
\def\r{\rho}
\def\s{\sigma}
\def\e{\epsilon}
  \def\scri{\mathcal{J}}
\def\cM{\mathcal{M}}
\def\tcM{\tilde{\mathcal{M}}}
\def\RR{\mathbb{R}}
\hyphenation{re-pa-ra-me-tri-za-tion}
\hyphenation{trans-for-ma-tions}
\begin{flushright}
IFT-UAM/CSIC-07-32\\
arXiv:YYMM.NNNNvV\\
\end{flushright}
\vspace{1cm}
\begin{center}
{\bf\Large   A comment on the matter-graviton coupling}\\
\vspace{.5cm}
{\bf Enrique \'Alvarez and Ant\'on F. Faedo}
\vspace{.3cm}
\vskip 0.4cm  
 
{\it  Instituto de F\'{\i}sica Te\'orica UAM/CSIC, C-XVI,
and  Departamento de F\'{\i}sica Te\'orica, C-XI,\\
  Universidad Aut\'onoma de Madrid 
  E-28049-Madrid, Spain }
\vskip 0.2cm
\vskip 1cm
\begin{abstract}
We point out a generic inconsistency of the coupling of ordinary gravity as described by general relativity 
with matter invariant only under unimodular diffeomorphisms (TDiffs), and some previously studied 
exceptions are pointed out. The most general Lagrangian invariant under TDiff up to dimension five
operators is determined, and consistency with existing observations is studied in some cases. 
\end{abstract}
\end{center}
\begin{quote}
\end{quote}
\newpage
\setcounter{page}{1}
\setcounter{footnote}{1}
\newpage

\section{Introduction}

There is a well known way of obtaining the  General Relativity Lagrangian 
which is associated with the name of Feynman \cite{Feynman}, although many other scientists have
contributed to it, starting with Kraichnan \cite{Kraichnan} \footnote{Some further references can be found in the review 
article \cite{Alvarez} or in the book by Ortin \cite{Ortin}}.
The idea is the following: if one starts from the Fierz-Pauli Lagrangian (which describes free spin two 
particles in Minkowski space)
\be
L_{FP}=\frac{1}{4}\pd_\m h^{\n\rho}\pd^\m h_{\n\rho}-
\frac{1}{2}\pd_\m h^{\m\rho}\pd_\n h^\n_\rho+\frac{1}{2}\pd^\m h\pd^\rho h_{\m\rho}-
\frac{1}{4}\pd_\m h \pd^\m h
\ee
where all indices are raised and lowered with the flat Minkowski metric; in particular:
\bea
&&h^{\m\n}\equiv \eta^{\m\a}\eta^{\n\b}h_{\a\b}\nonumber\\
&&h\equiv \eta^{\a\b}h_{\a\b}
\eea
it so happens that the equations of motion $D^{FP}_{\m\n}\equiv\frac{1}{2}\frac{\d S_{FP}}{\d h^{\m\n}}$
 are transverse, i.e.
\be
\pd^\m D^{FP}_{\m\n}=0
\ee
In order to coupling the graviton field $h_{\m\n}$ to an scalar field $\phi$, say, it is 
natural to try the coupling to the conserved energy-momentum tensor
(suitably symmetrized if needed, for example using  the Belinfante technique), that is
\be
L_I\equiv h^{\m\n} T_{\m\n}
\ee
But when this term is added to the matter Lagrangian in an freely falling inertial frame
\be
L^0_m=\frac{1}{2}\eta^{\m\n}\pd_\m\phi\pd_\n\phi -  V(\phi) 
\ee
the former energy-momentum tensor is no longer conserved, and the gravitational equation of motion
is inconsistent. This leads to a series of 
modifications that eventually end up in the Hilbert Lagrangian.
The quickest path to is probably Deser's, \cite{Deser} using a first order formalism. 
The aim of the present paper is to explore what room is left in this argument
for less symmetric nonlinear completions, notably the ones we dubbed TDiff, which are 
invariant under coordinate transformations whose jacobian enjoys unit determinant. 
These have been explored in \cite{AlvarezBGV}, where further references can be found.

\section{The linear approximation}

It is nevertheless clear that given a consistent theory (such as General 
Relativity itself) its linear part in any analytic expansion should be consistent as 
well (up to linear order). The object of our concern in the present paper will be the 
linear deviations from flat Minkowski space, i.e.
\be
g_{\m\n}=\eta_{\m\n}+\kappa h_{\m\n}
\ee
where $\eta_{\m\n}$ is the Minkowski metric, and $\kappa^2\equiv 8\pi G$.
This equation is taken to be an exact one; it can be looked at as the definition of $h_{\m\n}$.
\par
Now, it is a fact of life that $l_P\equiv \kappa\sqrt{\frac{ \hbar}{c^3}}$ has got 
dimensions of length, and that $M_P\equiv\frac{\sqrt{\hbar c}}{\kappa}$ enjoys
dimensions of mass. The value of Newton' s constant indicates that at the scale of terrestial experiments,
$M_P\sim 10^{19}GeV$. This means that the field $h_{\m\n}$ enjoys the proper canonical 
dimension (one) of a four-dimensional gauge field.
\par
The inverse metric is defined as a formal power series:
\be
g^{\m\n}=\eta^{\m\n}-\kappa h^{\m\n}+\kappa^2 h^\m_\s h^{\s\n}-\kappa^3 h^{\m\s} h_{\s\rho} h^{\rho\n} + o(\kappa^4)
\ee
Diffeomorphisms with infinitesimal parameter $\xi^\m$ act on the full metric as
\be
\d g_{\m\n}=\pounds(\xi)g_{\m\n}
\ee
whereas in terms of the fluctuations 
\be\label{variations}
\d h_{\m\n}=\xi^\rho \pd_\rho h_{\m\n} +\frac{1}{\kappa}\left(\pd_\m\xi_\n+\pd_\n\xi_\m\right)+h_{\m\a}\pd_\n\xi^\a +h_{\a\n}\pd_\m\xi^\a
\ee
This is, again, an exact formula, in the sense that there are no $\kappa$ corrections to it.
\par
The symmetry as above, without any restrictions, is the one corresponding to General Relativity, and in the present paper will be referred to simply as Diff.
When the vector $\xi^\a$
is restricted to
\be
\pd_\a\xi^\a=0
\ee
the symmetry is broken to what we call T(ransverse)Diff \cite{Blas}.
\par
The total action is then defined by
\be
S=\frac{1}{2}S_{h}+S_m
\ee
Here $S_h$ represents the purely gravitational sector, which is the most general 
Lorentz invariant dimension four operator that can be written with the field
$h_{\a\b}$, and its derivatives. It can be parametrized by a string of  
constants, namely the ones associated with the kinetic energy, $c_i$,$i=1\ldots 3$ and the ones associated to the potential energy for the fluctuations, 
which is the most general quartic potential in the fluctuations $h_{\a\b}$, namely, $\l_i$, $i=1\ldots 11$. 
The overall scale of the potential energy is related to  the {\em cosmological constant},
$\Lambda\equiv M_D^4$.
Cosmological observations seem to favor the tiny value $M_D\sim 10^{-3} eV$.
\bea\label{grav}
S_h&\equiv&\int d^4 x \left(\frac{1}{4}\pd_\m h^{\n\rho}\pd^\m h_{\n\rho}-
\frac{c_1}{2}\pd_\m h^{\m\rho}\pd_\n h^\n_\rho+c_2 \frac{1}{2}\pd^\m h\pd^\rho h_{\m\rho}-
c_3\frac{1}{4}\pd_\m h \pd^\m h +\right.\nonumber\\
&&M_D^4 \left(1+\frac{1}{2M_P} \l_1 h +\frac{1}{8M_P^2} \left(\l_2 h^2-2 \l_3 h_{\a\b}h^{\a\b}\right)+\right.\nonumber\\
&&\left.\left.\frac{1}{48 M_P^3 }\left(\l_4 h^3+8 \l_5 h_{\m\n} h^{\n\rho} h_\rho^\m - 6 \l_6 h h_{\a\b}h^{\a\b} \right)+\right.\right.\nonumber\\
&&\left.\left.\frac{1}{384 M_P^4}\left(\l_7 h^4-12 \l_8 h^2 h_{\a\b}h^{\a\b}+32 \l_9 h h_{\a\b}h^{\b\gamma}h_\gamma^\a -\right.\right.\right.\nonumber\\
&&\left.\left.\left.48 \l_{10}h_{\a\b}h^{\b\gamma}
h_{\gamma\delta} h^{\delta\a}+12 \l_{11}\left(h_{\a\b}h^{\a\b}\right)^2\right)\right)
\right)
\eea
Let us remark, first of all, that the structure of the symmetry transformations 
of both Diff and TDiff is such that terms in the Lagrangian of $O(M_P^{-n})$ are related
to terns of both $O(M_P^{-n})$ as well as, because of the abelian part, terms of $O(M_P^{-n+1})$. This means that 
in the variations of the kinetic energy part we can keep only the
piece in $M_P$ , since the other part ($O(M_P^0)$) of the variation 
should cancel with the $M_P$ contribution to the kinetic operators of order
$O(\frac{1}{M_P})$, which we have not considered.
The only piece we can consistently consider there is then the Fierz-Pauli abelian part given by
\be
\d h_{\m\n}=M_p\left(\pd_\m\xi_\n+\pd_\n\xi_\m\right)
\ee
The situation is different, however, in the potential energy piece. In order to cancel 
the Fierz-Pauli variation of the $O(M_P^{-2})$ term, it is necessary to 
consider the $O(M_P^0)$ variation of the $O(\frac{1}{M_P})$ term. This means that the full action
 is invariant under the full variation (\ref{variations}) up to dimension five operators
(which means $O(\frac{1}{M_P})$ in the kinetic energy part, and $O(\frac{1}{M_P^{-5}})$ in the potential
energy piece).
\par
Under those provisos, TDiff needs that
\bea
&&\l_1=\l_3=\l_5=\l_{10}\nonumber\\
&&\l_2=\l_6=\l_9=\l_{11}\nonumber\\
&&\l_4=\l_8
\eea
The most general TDiff invariant potential depends on four arbitrary parameters.
In some studies it is frequent to restrict the gravitational equations of motion 
to the linear approximation; this means quadratic terms in the gravitational lagrangian.
From a field theoretical viewpoint there is no reason lo leave away any {\em relevant}
(in the renormalization group sense) operators.
\bi
\item Let us first consider the case $\l_i=0\quad \forall i$.
This corresponds  to vanishing cosmological constant in general relativity.
First of all, TDiff enforces $c_1=1$.
\par
Besides, there are two exceptional values, namely $c_i=1, \forall i $, when TDiff is 
enhanced to full Diff . 
This is the only combination for which the wave operator is transverse.
\bea
\frac{1}{2}\frac{\d S_h}{\d h^{\a\b}}&\equiv& D^h_{\a\b}=-\frac{1}{4}\Box h_{\a\b}+
\frac{c_1}{4}\left(\pd_\rho\pd_\a h_\b^\rho+\pd_\rho\pd_\b h^\rho_\a\right)-
\frac{c_2}{4}\left(\eta_{\a\b}\pd_\m\pd_\n h^{\m\n}+\pd_\a\pd_\b h\right)+\nonumber\\
&&+\frac{c_3}{4}\Box h\, \eta_{\a\b}
\eea
Indeed
\be
\pd^\a D^h_{\a\b}=\frac{c_1-1}{4}\Box\pd^\a h_{\a\b}+
\frac{c_1-c_2}{4}\pd_\b \pd_\rho\pd_\sigma h^{\rho\sigma}+
\frac{c_3-c_2}{4}\Box \pd_\b h 
\ee
By the way, it is worth noticing that the metric condition
\be
\nabla_\m g_{\a\b}=0
\ee
is {\em identically} satisfied to $o(\kappa)$ and poses no restriction on $h_{\m\n}$.
\par
\item The other remarkable value is $c_1=1,\, c_2=\frac{1}{2},\, c_3=\frac{3}{8}$ where the symmetry is enhanced 
with a Weyl invariance, denoted by WTDiff, and the 
wave operator is traceless in the absence of cosmological constant. To be specific,
\be
\eta^{\m\n}D^h_{\m\n}=\left(c_3-\frac{c_2+1}{4}\right)\Box h +\left(c_2-\frac{c_1}{2}\right)
\pd_\m\pd_\n h^{\m\n}
\ee
The analysis in \cite{AlvarezBGV} shows that these two are the only instances 
where only spin two is present, with no scalar contamination.
\item Let us now consider the effect of $\l_i\neq 0$. First of all, Diff invariance is recovered when $\l_i=1\quad \forall i$. Curiously enough, as such, and in the quadratic approximation,
the term
\be
m_1^2\equiv\frac{M_D^4}{M_p^2}\l_3
\ee
as well as 
\be
m_2^2\equiv\frac{M_D^4}{2 M_p^2}\l_2
\ee
do have the interpretation of {\em masses}. 
\par
Only when the background around which we perturb is not flat, but a constant curvature space, with metric
$\bar{g}_{\m\n}$,
do these paramerers recover the meaning of cosmological constant. In that case it is mandatory to
substitute all derivatives by background covariant derivatives, i.e.,
\be
\pd_\a h_{\m\n}\rightarrow\bar{\nabla}_\a h_{\m\n}
\ee
and to raise and lower indices using the corresponding background metric:
\be
h^{\m\n}\equiv \bar{g}^{\m\a}\bar{g}^{\n\b}h_{\a\b}
\ee
\par
\item Let us study now consistency of the coupling, which was the main motivation
of this work. The matter Lagrangian has been denoted by $S_m$.
Up to dimension five operators  for a scalar field, which in a free falling locally inertial 
reference system has lagrangian
\be
L^0_m=\frac{1}{2}\eta^{\m\n}\pd_\m\phi\pd_\n\phi -  V(\phi) 
\ee
assuming a $\mathbb{Z}_2$ symmetry
\be
\phi\rightarrow -\phi
\ee
the allowed matter operators when a gravitational field is present can be parameterized 
by three constants, $\m_1\ldots \m_3$:
\be\label{matt}
L_m=\frac{1}{2}\eta^{\m\n}\pd_\m\phi\pd_\n\phi -V(\phi)+ 
\frac{1}{M_P}\left(-\frac{\m_1}{2}  h^{\m\n}\pd_\m\phi\pd_\n\phi+ 
\m_2 \frac{1}{4} h\eta^{\m\n}\pd_\m\phi\pd_\n\phi-  
\m_3 \frac{h}{2}V(\phi)\right)
\ee
Remember that the variation of a scalar field is
\be
\d\phi=\xi^\m \pd_\m \phi
\ee
In order to enjoy TDiff invariance, it is necessary that $\m_1=1$.
Diff invariance needs in addition that   $\m_2=\m_3=1$.
The matter equations of motion are
\be
\frac{\d S_m}{\d \phi}=-\Box \phi- V^{\prime}(\phi)+
\frac{1}{M_p}\left(\m_1\pd_\a\left(h^{\a\b}\pd_\b\phi\right)-
\frac{\m_2}{2}\pd_\a\left(h \pd^\a\phi\right)-\m_3 \frac{h}{2} V^{\prime}(\phi)\right)
\ee
and the gravitational equations
\be\label{graveq}
\frac{\d\left(\frac{S_h}{2} + S_m\right)}{\d h_{\m\n}}=D^h_{\m\n}-
\frac{1}{M_p}\left(\frac{1}{2}\m_1\pd_\m\phi\pd_\n\phi+\frac{1}{2}\left(
\m_3 V(\phi)-\frac{\m_2}{2}(\pd_\a\phi)^2\right)\eta_{\m\n}\right)
\ee
There is generically no problem of consistency, except in the two exceptional  cases.
\par
First of all, when $S_h$ has extended Diff symmetry, and consequently a transverse
 wave operator, this forces $S_m$ to have the same Diff symmetry through the linear 
expansion of the $\sqrt{|g|}$ term; otherwise consistency of the coupling enforces 
extra condictions on the matter, a very 
weird situation indeed (i.e. Bianchi identities are still valid on the gravitational side, so that by consistency the same identities must hold true on the matter side as well). Nevertheless, it is not fully devoid of interest to 
study the situations in which 
there are exceptions to this rule; this we did in a previous work \cite{AlvarezF}.
\par
\item When there is WTDiff symmetry, it is clear that the matter lagrangian should also be 
scale invariant 
in order for the corresponding energy-momentum to be traceless.  In our example, 
this corresponds to $\m_1= 2 \m_2$ and $\m_3=0$.
\par
\item There are models in which Diff invariance in the matter sector is reached using in the volume element  some other scalar density, such as the square root of the determinant of a matrix built 
out of fields and their derivatives (as in the very interesting ones proposed in
\cite{Guendelman}) instead of the $\sqrt{|g|}$ term implicit in the metric volume 
element. The tensor that appears as the source of gravity in Einstein's equations is covariantly conserved thanks to the equations of motion of the fields in the scalar density\footnote{Although its flat limit seems to be different from the canonical energy-momentum tensor.}. Of course that tensor is not the usual energy-momentum tensor of General Relativity, which now is not conserved. The reason is that in order for the Rosenfeld energy-momentum tensor to be equivalent to the canonical 
Belinfante one what is needed is not only Diff invariance, but also the standard metric 
volume element \cite{Babak}. This topic seems worthy of some further investigation.
\ei

\section{Observational constraints}

In this section we will outline the way in which one can constraint the space of paramaters of the linearized
theory (i.e. $c_i$, $\m_i$ and $\l_i$) using experimental results on deviations from 
Newton's inverse square law. For simplicity we will illustrate with a very particular example so no definite conclusions
can be drawn concerning the viability of this kind of models. 
\par
Detailed computations of the propagators can be found in \cite{AlvarezBGV}, where the authors considered
a gravitational Lagrangian (\ref{grav}) with all $\l_i=0$ except $m_2^2\equiv\frac{M_D^4}{2 M_p^2}\l_2$
which has the interpretation of a mass for the scalar part of the graviton, present generically in this
kind of models with TDiff invariance.
It turns out that for a conserved energy-momentum tensor coupled to gravity in the form
\be\label{coup}
L_I=\frac{1}{2}h_{\m\n}\left(\kappa_1T^{\m\n}+\kappa_2\eta^{\m\n}T\right)
\ee
then in momentum space the interaction is
\be
L_I=\kappa_1^2\left[T^{*}_{\m\n}T^{\m\n}-\frac{1}{2}|T|^2\right]\frac{1}{k^2}-\left(\kappa_2+\frac{1-c_2}{2}\kappa_1\right)^2\frac{|T|^2}{\Delta ck^2-m_2^2}
\ee
where we have defined
\be
\Delta c=c_3-\frac{1}{2}+c_2-\frac{3}{2}c_2^2
\ee
with the constraint $\Delta c <0$ because of unitarity \cite{AlvarezBGV}. The first term corresponds to the usual spin 2 exchange 
while the second one is an additional massive scalar interaction. Let us turn our attention to a particular example, namely the
matter Lagrangian (\ref{matt}). Unfortunately the corresponding energy momentum tensor is not conserved. However, a conserved
tensor can be defined as
\bea
\Theta_{\m\n}&\equiv &T_{\m\n}-\frac{1}{2}\eta_{\m\n}\left(\frac{1-\m_2}{2}(\pd_\r\phi)^2+(\m_3-1)V(\phi)\right)+ O(\frac{1}{M_p})=\nonumber\\
&=&\frac{1}{2}\pd_\m\phi\pd_\n\phi-\frac{1}{2}\eta_{\m\n}\left(\frac{1}{2}(\pd_\r\phi)^2-V(\phi)\right)+ O(\frac{1}{M_p})
\eea
In the particular case that $\m_3=2\m_2-1$ (which includes the Diff invariant Lagrangian) our energy-momentum tensor 
can be written in terms of the new one and its trace in such a way that the coupling $M_p^{-1}h_{\m\n}T^{\m\n}$ is 
of the form (\ref{coup}) with
\bea
&&\kappa_1=\frac{2}{M_p}\nonumber\\
&&\kappa_2=\frac{\m_2-1}{M_p}
\eea
Now we can apply directly the preceeding results and study experimental constraints to this model. 
The exchange of additional massive scalar degrees of freedom produces a Yukawa like potential which 
is usually parametrized as \cite{Nelson}
\be
V(r)\sim\frac{1}{r}\left(1+\a e^{-\frac{r}{\l}}\right)
\ee
The parameter $\a$ is then the ratio between the spin 2 and the scalar couplings, in our particular case
\be\label{alfa}
\a=-\frac{\left(\kappa_2+\frac{1-c_2}{2}\kappa_1\right)^2}{\Delta c\,\,\kappa_1^2}=-\frac{\left(\m_2-c_2\right)^2}{4\,\Delta c}
\ee
While $\l$ gives the range of the interaction, or equvalently the mass of the scalar exchanged
\be\label{landa}
\l^2=\frac{\Delta c}{m_2^2}
\ee
Notice that one has to impose $m_2^2<0$ since as we have said absence of ghosts requires $\Delta c <0$. 
\par
\begin{figure}[h]
\centering
\includegraphics[scale=0.6,angle=270]{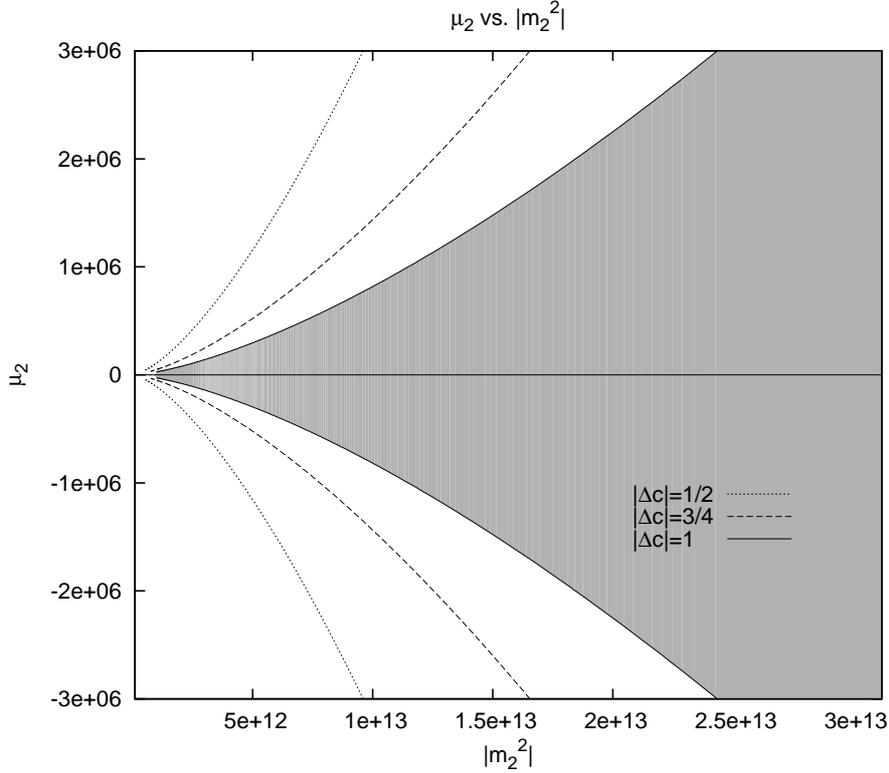}
\caption{\label{m2.n2}\small{The shadowed region shows experimentally allowed values for $|m_2^2|$ (in $m^{-2}$)
and $\m_2$ for given values of $c_2$ and $c_3$, expressed in terms of $\Delta c$, and in the range $\l \in$
$10^{-9}$m-$10^{-6}$m.}}
\end{figure}

There are important constraints on the strength of hypothetical Yukawa interactions for a wide range
of $\l$. Through (\ref{alfa}) 
and (\ref{landa}) it is then possible to constraint the space of parameters of the linearized theory.
We will use figures 4, 5 and 9 of reference \cite{Nelson}, which show regions allowed and excluded for
$\a$ corresponding to $\l$ in the ranges $10^{-9}$m-$10^{-6}$m, $10^{-6}$m-$10^{-2}$m and $10^{-2}$m-$10^{14}$m respectively.
Since we are just interested in general behaviours and not in accurate results we will approximate
 the experimental curves by straight lines. The original plots are in logarithmic scale so we have
experimentally allowed regions of the form
\be\label{curve}
|\a|<k\,\l^a
\ee
We just have to substitute this expresion into (\ref{alfa}) to get bounds for our parameters.
There are however four parameters to play with ($\m_2$, $m_2^2$, $c_2$ and $c_3$). First, it is interesting
to see the order of magnitude for the mass once we fix the values of $c_2$ and $c_3$. The result is ploted
in Fig.\ref{m2.n2}. It can be seen that greater values for the mass are favoured, being the lower bound around
$|m_2^2|\sim 5\times 10^{11}\mathrm{m}^{-2}\sim 0.02 \,\,\mathrm{eV}^2$, and that the allowed region rapidly decreases with $|\Delta c|$.

\begin{figure}[h]
\centering
\includegraphics[scale=0.5,angle=270]{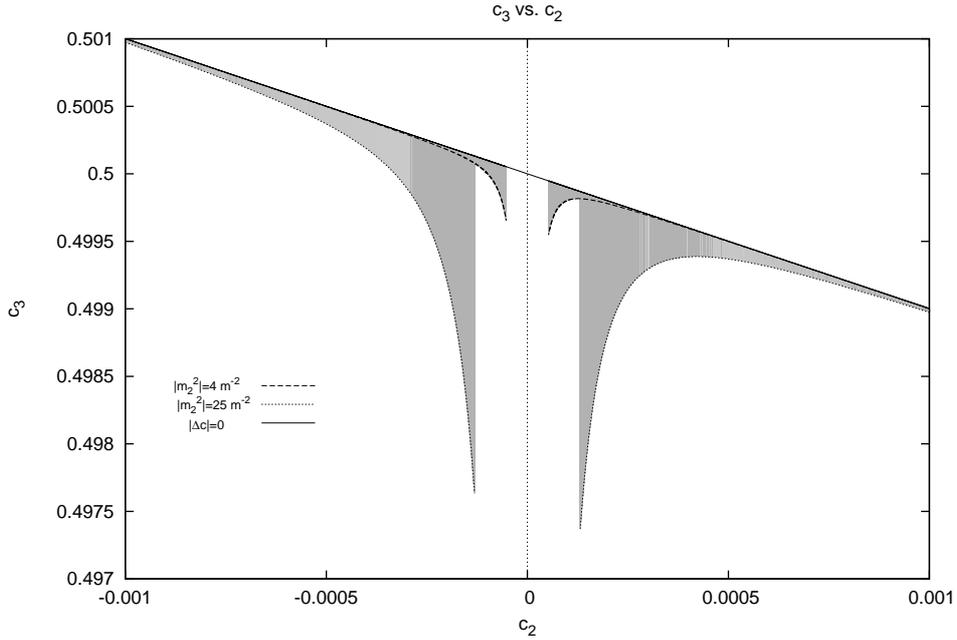}
\caption{\label{c3.c2.1}\small{Experimentally allowed region in the plane ($c_2$,$c_3$), for a couple 
of values of the mass, in the range $\l\in10^{-6}$m-$10^{-2}$m. The plot is restricted to the zone where the
curve appreciably deviates from the parabola $\Delta c=0$.}}
\end{figure}

\begin{figure}[!htbp]
\centering
\includegraphics[scale=0.4,angle=270]{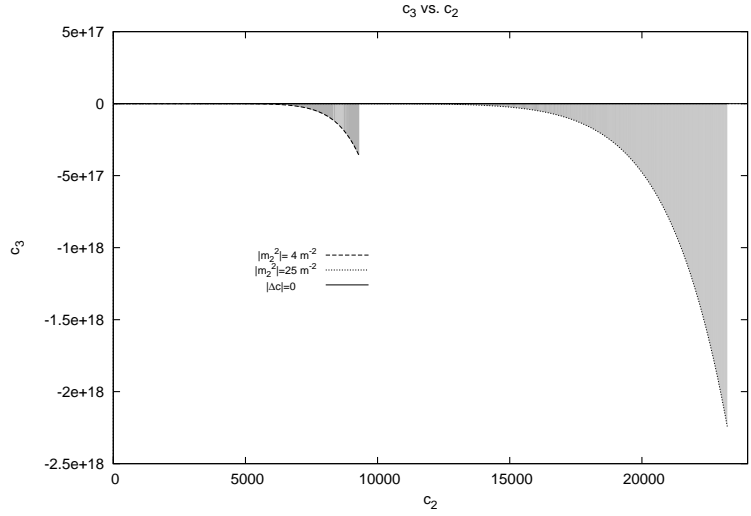}
\caption{\label{c3.c2.2}\small{Experimentally allowed region in the plane ($c_2$,$c_3$), for a couple 
of values of the mass, in the range $\l\in10^{-2}$m-$10^{14}$m. We only show the positive $c_2$ branch.
The parabola is indistinguisable from the $c_2$ axis.}}
\end{figure}

\begin{figure}[!htbp]
\centering
\includegraphics[scale=0.4,angle=270]{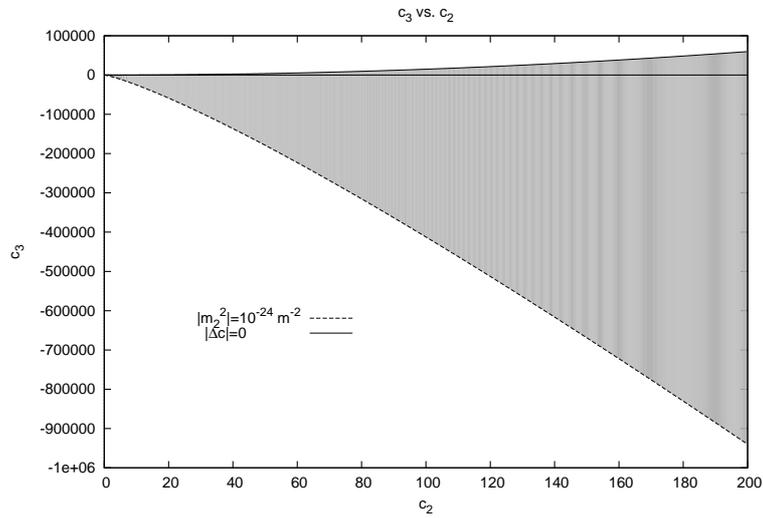}
\caption{\label{c3.c2.3}\small{Experimentally allowed region in the plane ($c_2$,$c_3$) and in the range 
$\l\in10^{-2}$m-$10^{14}$m for a very tiny mass. That allows us to see the parabola, which was hidden
in the previous figure. We only show the zone closest to the $c_3$ axis.}}
\end{figure}

Another possibility is to fix $|m_2^2|$ and $\m_2$ and see, in the plane ($c_2$,$c_3$), how far from Diff
invariance (which corresponds to $c_2=c_3=1$) can we move away. Remember that we also have to 
take into account the restriction $\Delta c<0$. For the first range, $\l\in10^{-9}$m-$10^{-6}$m there is no hope of seeing
an experimental curve that appreciably deviates from the parabola $\Delta c=0$ because of (\ref{landa}) and the tiny
values of $\l$. Increasing $|m_2^2|$ does increase the allowed region, which is between both curves, but does not 
produce a plot in which the curves are visibly separate. It can be understood if we realise that increasing the mass
also increases $c_3$ on the parabola through $c_2$ and (\ref{alfa})-(\ref{landa}). An approximate definition of the
separation could be 
\be
Sep\sim\frac{(c_3)_{par}-(c_3)_{cur}}{(c_3)_{par}}=\frac{\l^2|m_2^2|}{\frac{1}{2}-c_2(\l,m_2)+\frac{3}{2}c_2(\l,m_2)^2}
\ee
where $(c_3)_{par}$ and $(c_3)_{cur}$ means the value of $c_3$ on the parabola and the experimental curve respectively.
For the first experimental range one has $Sep<<1$ independently of the mass and in the whole interval.
The other two cases do not have that property, which is of course related also with the particular
values of $k$ and $a$ in (\ref{curve}). Examples of resulting plots are Figs.\ref{c3.c2.1},$\,$\ref{c3.c2.2} and \ref{c3.c2.3}.

Once again the experiment prefers greater values for the mass. All the plots have $\m_2=0$, other values just
move the experimental curve along the parabola, but the qualitative result remains unchanged.

\newpage
\section{Conclusions}

In this paper we have studied at the linearized level the viability of gravity models with a restricted symmetry, both from the theoretical and observational points of view. While the existing observational constraints on additional Yukawa like gravitatory interactions do not seem to be a major obstacle, a consistency problem has been identified. At the non linear level it appeared as an integrability condition on Einstein's equations  \cite{AlvarezF}. Here we turned our attention to the linear level in order to see if the problem could be avoided, and if so in what type of more clever non linear completions. 
The main conclusion is that it is not generically possible to couple matter  (i.e., 
with an arbitrary equation of state) to gravitation in such a way
that this coupling has a restricted symmetry only (what has been called TDiff) 
whereas the purely gravitational sector enjoys a higher symmetry, namely
the standard Diff invariance, or an additional Weyl symmetry (WTDiff). That this is possible in some restricted cases has been already found in a previous paper  \cite{AlvarezF}.
\par
The condition for arbitrary TDiff matter to be able to couple to Diff gravity
 without restrictions can be stated somewhat more formally by saying that the Rosenfeld 
(metric) energy-momentum tensor has got to be equivalent to the Belinfante canonical form.
\par
On the other hand, there is a widespread {\em urban legend} asserting that unimodular theories are 
equivalent to General Relativity with a cosmological constant. Specific
calculations both here and in our previous paper \cite{AlvarezF} have proven it to be 
groundless.  It is a fact that in some TDiff models there is no exponential
expansion at all which is well known to be the benchmark of a (positive) cosmological constant
in General Relativity. Therefore, that models provide a counterexample to the statement above.
\par
Nevertheless, as with all legends, there is some partial truth in it.
The equations of motion of the example in \cite{AlvarezF} correspond to $c_1=c_2=c_3=\m_1=1$ and 
$\m_2=\m_3=\l_i=0$, that is
\be\label{T}
D^{FP}_{\a\b}=\frac{1}{M_P}\pd_\a\phi\pd_\b\phi
\ee
whereas the linear equations of General Relativity with cosmological constant read:
\be\label{GR}
D^{FP}_{\a\b}+\frac{\l M_P^3}{4}\eta_{\a\b}
=\frac{1}{M_P}\left(\pd_\a\phi\pd_\b\phi+
\frac{1}{2}\left(V(\phi)-\frac{1}{2}(\pd_\rho\phi)^2\right)\eta_{\a\b}\right)
\ee
Now, the equations (\ref{T}) are inconsistent as such, in the sense that only a subsector
of the theory, namely, the one that obeys
\be\label{IC}
V-\frac{1}{2}(\pd_\rho\phi)^2= C
\ee
 can be coupled to gravitation. There are many sectors of matter in 
a freely falling 
inertial system that do not obey this \footnote{This  physically means that 
the pressure vanishes, (for a perfect fluid the generally covariant lagrangian can be 
identified with the physical pressure cf.\cite{Hawking}), i.e. that the matter is 
what cosmologists
call {\em dust}}restriction. Actually, together with energy conservation, the 
aforementioned equation implies that both the kinetic and potential energy ought
to be constant:
\bea
&& 2 V(\phi)= E + C\nonumber\\
&& (\pd_\rho\phi)^2 = E - C
\eea 
It is clear that for a scalar field in flat space most initial conditions lead to
configurations that violate those equations. 
This would mean that an inconsistency
would show  up once a gravitational field is turned on, however weak. 
More formally, something very strange should happen when changing the reference frame
from an inertial (freely falling one) to another in which a gravitational field is
present.
\par 
Those are the reasons what we say 
that the coupling is generically inconsistent. Let us accept nevertheless, for the 
sake of the argument, that physics is so restricted. Then a glance at the equations (\ref{GR}) 
of General Relativity shows that they are indeed equivalent to (\ref{T}) {\em provided}
we identify
\be
\l\equiv \frac{2 C}{M_P^4}
\ee
But this is only due to our choice of the arbitrary constants, and under the asumption that the coupled sector is only the one that obeys (\ref{IC}), a deeply misterious condition from a General Relativistic perspective. In the general TDiff case the analogous condition to (\ref{IC}) is
\be
-\frac{\m_2-1}{2}(\pd_\rho\phi)^2+(\m_3-1)V(\phi)=C
\ee
and the system is not equivalent to General Relativity with a cosmological constant.
\section*{Acknowledgments}
    
A.F.F. would like to thank Irene Amado for her infinite patience and help with the figures. 
This work has been partially supported by the European Commission (HPRN-CT-200-00148) and 
by FPA2003-04597 (DGI del MCyT, Spain), as well as Proyecto HEPHACOS (CAM); P-ESP-00346.      
A.F.F has been supported by a MEC grant, AP-2004-0921. 
               
\end{document}